# COMBINATION OF FREQUENCY SHIFT AND IMPEDANCE-BASED METHOD FOR ROBUST TEMPERATURE SENSING USING PIEZOCERAMIC DEVICES FOR SHM

Emmanuel LIZE[1], Charles HUDIN[1], Nicolas GUENARD[1], Marc REBILLAT[2], Nazih MECHBAL[2], Christian BOLZMACHER[1]

[1] CEA, LIST, Sensorial and Ambient Interfaces Laboratory, 91191 - Gif-sur-Yvette CEDEX, France;
[2] Processes and Engineering in Mechanics and Materials Laboratory (PIMM, UMR CNRS 8006, Arts et Métiers ParisTech (ENSAM)), 151, Boulevard de l'Hôpital, Paris, F-75013, France; emmanuel.lize@cea.fr

**Key words:** guided waves (lamb waves), Piezoelectric sensors, composite, temperature, impedance, frequency-shift, SHM.

**Abstract**
*The influence of temperature is a major problem in Structural Health Monitoring diagnosis using guided waves. In this article, two methods for temperature compensation are used to evaluate the temperature of a structure monitored by Piezoceramic Transducers (PZT). The first one consists in a linear regression of the static capacity of the PZT (computed with the electromechanical impedance) with the temperature. It allows one to determine temperature of a PZT from any impedance measurement. The second method is based on the Modes Frequency Shift (MFS) of the frequency response function (FRF) calculated in a pitch-catch mode with an exponential sweep signal. A linear regression between the MFS and the temperature is settled for most modes and allow the estimation of temperature between two PZT. Experiments on a composite plate monitored by 5 PZT patches showed that a $\pm 2°C$ variation can easily be identified with both methods. Combined together, those two methods could replace actual temperature sensors and bring an efficient improvement for temperature compensation in applications where the temperature variation is heterogeneous on a structure.*

## 1 INTRODUCTION

Structural Health Monitoring (SHM) is a field that has been continuously evolving for the last 20 years. The revolution of new technologies and improvements in computer science leads to great innovations on existing techniques, such as signal processing, that allow one to create more complex methods and bring more accurate results in SHM. However, some important parameters are still difficult to handle, including the influence of environmental conditions and the complexity of materials of monitored structures. The aerospace industry is specifically concerned by these technological improvements since composite materials tend to replace aluminum and the maintenance becomes quite expensive.

Damage detection using guided waves is one of the most used methods in SHM. The ultrasonic wave propagation in a structure is altered by the material properties, the geometry and the presence of damages. Among all the methods employed to trigger and catch guided waves, the use of piezoelectric transducer (PZT) is one of the cheapest and easiest-to-settle.

The study of electromechanical signature of PZT devices has gained importance in the SHM community [1]–[3]. It consists in measuring the impedance of PZT coupled with a structure under test. This method can be used as a damage-detection method and to evaluate a PZT boundary condition or health degradation. It has shown very good results under laboratory conditions and has already been deployed under real-life conditions. In an aeronautical context, the influence of temperature variations on the impedance of surface-mounted PZT is not negligible and has been the focus of several research teams. Experimental [4]–[6] as well as numerical studies [7], [8] pointed out that the static capacity of a PZT increases, and the modal frequencies are lowered when the temperature rises [9]–[11]. Those observations have led to temperature compensation methods, and efforts have been made to distinguish modifications attributable to temperature from modifications due to damage [10]–[13]. All the studies are efficient in proving the sensibility of PZT to temperature, but to the authors' knowledge, very few works have been carried out in order to use the PZT as a temperature sensor using static capacity. Ilg & al. [14] demonstrated that it is possible to use an air ultrasound transducer to estimate the air temperature with a $\pm 4.5°C$ accuracy using the static capacity of PZT at 1kHz. Yet, this study didn't show results of surface mounted PZT.

Since wave propagation in structures is used in SHM, its sensibility to temperature variation has been investigated in numerous studies [15], [16]. It has been demonstrated that changes of properties and geometry of the structure being caused by temperature variations are one of the environmental parameters that reduces the most significantly the probability of detection of damages in a structure [17]. Many studies exist on homogenous structures such as aluminum plates, but fewer work has been applied to composite structures as Carbon Fiber Reinforced Plastic (CFRP). However, some effective temperature-compensation models already exist. Those methods are mostly based on time domain measures [18] or their interpretation in the frequency domain. For example, Crowford & al. [19], [20] introduced the Optimal Baseline Selection (OBS) and the Baseline Signal Stretch (BSS) in the time domain. Clarke et al. [21] used the same methods in the frequency domain. The combination of those 2 techniques is effective and enables to have smaller baselines. More complex algorithms have been developed using this combination [22]. The main influence of temperature on the frequency response function (FRF) resides in a variation of the modal damping and a frequency shift of the modal resonances. This observation is used for temperature compensation (BSS in the frequency domain) but has never been used to propose a way to determine the temperature of the structure between an actuator-sensor couple.

This article is organized as follows. The static capacity and the modal frequency-shift temperature identification methods are first described. Then the experimental set-up used to study the influence of temperature and the effectiveness of the proposed identification methods is presented. Finally, experimental results are displayed.

## 2 STATIC CAPACITY AND MODAL FREQUENCY SHIFT

This section defines the static capacity of a PZT mounted on a structure and how it can be used to determine temperature. It also presents the frequency response function (FRF) obtained with an exponential sweep and how the frequency shift of the observed modes can be used to evaluate the temperature. Eventually, two indicators are described: the coefficient of determination and the error of temperature estimation.

## 2.1 Static capacity computation

A PZT mounted on a structure can be qualified by its electro-mechanical signature. The impedance $Z(f)$ and its inverse, the admittance $Y(f)$, are defined as the transfer function between voltage $V(f)$ applied to the PZT element and the resulting current $I(f)$ The equivalent capacity at a given frequency $C(f)$ is defined as the absolute value of the admittance divided by $2\pi f$. The variable $f$ denotes the frequency.

$$C(f) = \left|\frac{Y(f)}{2\pi f}\right| \text{ where } Y(f) = \frac{1}{Z(f)} = \frac{I(f)}{V(f)} \tag{1}$$

The static capacity is the mean of the equivalent capacity computed on a range of frequencies from $f_a$ to $f_b$:

$$C_s = \frac{1}{f_b - f_a}\int_{f_a}^{f_b} C(f)df \tag{2}$$

## 2.2 Estimation of the temperature using static capacity

Variations of temperature have a non-negligible effect on material properties of the structure under test, the PZT and the bounding. As a matter of fact, the static capacity of a PZT increases as the temperature rises and decreases as the temperature drops.

Experiments on a CFRP plate subject to temperature variation cycles between 0 and 80°C allow us to propose a linear regression of the static capacity with temperature:

$$C_s(\theta) = \alpha_C \times \theta + C_s^0 \tag{3}$$

The variables $\alpha_C$ and $C_s^0$ are respectively the temperature variation coefficient and the static capacity at 0°C. They are determined using experimental data acquisition of static capacity measurements. The variation of static capacity $\Delta C_s(\theta)$ is also defined as:

$$\Delta C_s(\theta) = C_s(\theta) - C_s^0 = \alpha_C \times \theta$$

Therefore, an estimation of the temperature of a PZT mounted on a structure $\hat{\theta}_{estim}$ can be calculated with the variation of static capacity for any measurement $\Delta C_{s,mes}$:

$$\hat{\theta}_{estim} = \frac{\Delta C_{s,mes}}{\alpha_C} \tag{4}$$

## 2.3 FRF using an exponential sweep signal

The exponential sweep signal is used to excite all the modes of the structure under test within a programed range of frequencies. Compared to other excitation signals composed of multiple sines (random multisine, linear sweep), the exponential sweep signal is more robust to non-linearities and less sensitive to distributed noises [23]. The FRF is the transfer function $H(\omega)$ which corresponds to the cross power spectral density of the signal recorded by the PZT sensor and the signal emitted by the PZT actuator divided by the power spectral density of the signal emitted by the PZT actuator. It is equivalent to the following equation:

$$H(\omega) = \frac{V(\omega) \times U(\omega)}{|U(\omega)|^2} \tag{5}$$

where $V(\omega)$ and $U(\omega)$ are the Fourier transform of the signal recorded by the PZT sensor and the signal emitted by the PZT actuator. The modes $r$ refer to the peaks observed in the FRF. They are defined by their pulsation at temperature $\theta$ by $\omega_r(\theta)$ in Hertz.

## 2.4 Influence of temperature on the FRF

The modal-frequency-shift (MFS) is the frequency variation of modes observed when the temperature changes. Its variation $\Delta\omega_r$ is estimated by

$$\Delta\omega_r(\theta) = \omega_r(\theta) - \omega_r^0 \tag{6}$$

where $\omega_r^0$ is the pulsation of the mode $r$ at 0°C.

In previous studies using FRF it has been noticed that the MFS towards lower frequencies increases for higher frequency and for rising temperature. For the first experiments, two main assumptions have been made concerning the MFS evolution:
  i. The MFS of each mode is linear with temperature
  ii. No regression model is applied regarding the MFS variation with frequency.

The thermal drift coefficient $\alpha_r$ of a given mode $r$ satisfies the relation

$$\Delta\omega_r(\theta) = \alpha_r \times \theta \tag{7}$$

It is important to notice that $\alpha_r$ corresponds to one specific mode. $\alpha_r$ and $\omega_r^0$ are determined using experimental data of FRF acquisitions.

## 2.5 Temperature estimation using the FRF

Using Eq.(7), the temperature of any FRF measurement can be estimated from the MFS measured for a given mode $\Delta\omega_{r,mes}$:

$$\hat{\theta}_{r,estim} = \frac{\Delta\omega_{r,mes}}{\alpha_r} \tag{8}$$

The use of a broadband signal allows the exploitation of several modes. It leads to a more robust result of the estimated temperature.

For $N$ considered modes, the mean of the estimated temperature $\overline{\theta_{estim}}$ is computed iteratively by minimizing the standard deviation $\sigma$, as described below. The final value of the mean calculated corresponds to the estimated temperature $\hat{\theta}_{estim}$.

Algorithm 1: Temperature estimation with $N$ modes.

Calculate: $\hat{\theta}_{estim} = \bar{\theta}_{estim} = \frac{\sum_{r=1}^{r=N}\hat{\theta}_{r,estim}}{N}$ and $\sigma = \sqrt{\frac{\sum_{r=1}^{r=N}[\hat{\theta}_{estim}(r) - \bar{\theta}_{estim}]}{N}}$

while $\exists \hat{\theta}_{r,estim} \notin [\bar{\theta}_{estim} \mp \sigma]$
    $Do$ for each $r$, delete $r$ if $\hat{\theta}_{r,estim} \notin [\bar{\theta}_{estim} \mp \sigma]$
    compute $N, \bar{\theta}_{estim}$ and $\sigma$ with the remaining $r$
end while
$\hat{\theta}_{estim} = \bar{\theta}_{estim}$

## 2.6 Indicators

The coefficient of determination $r^2$ is used as an indicator for linear regression verification. It is between 0 (erroneous linear regression) and 1 (perfect fit).

The error of estimation $\varepsilon_\theta$ is the absolute value of the difference between the estimated temperature $\theta_{estim}$ and the measured temperature $\theta_{meas}$:

$$\varepsilon_\theta = \theta_{meas} - \hat{\theta}_{estim} \tag{9}$$

## 3 EXPERIMENTS AND RESULTS

This section describes the setup used for the experiments and results obtained for temperature estimation with both methods (static capacity and MFS).

### 3.1 Structure under test and protocols

The testing was performed on a $400\ mm \times 300\ mm$ CFRP plate of 4 plies $[0/-45/45/0]$. Each ply is $0.28\ mm$ thick. This structure is instrumented with 5 Noliac NCE51 piezoceramics transducers with a diameter of $20\ mm$, a thickness of $0.5\ mm$ and a Curie temperature of $360°C$. Each PZT can be used as an actuator or a sensor. All the PZT were mounted on the structure with the glue Redux 322. According to the manufacturer data sheet, this glue can be used between $-55°C$ and $200°C$.

Impedance and FRF measurements were done at the same time. The CFRP plate was placed in an environmental chamber where the temperature could be controlled between $-60°C$ and $120°C$ (WEISS WKL100). A type K thermocouple was placed on the plate and acquired with a NI 9213 data acquisition card. A multiplexor AGILENT L4421 was used to switch PZT actuator/sensor or Impedance/FRF measurement. The acquisition system was monitored using Matlab. Impedance measurements were realized with a Hioki IM3570 impedance analyzer (see Figure 1.a). At each temperature of a cycle, the impedance measurement of PZTs was acquired three times with 800 points between $20\ kHz$ and $60\ kHz$. These measurements were then averaged in the frequency domain in order to increase the signal to noise ratio (SNR).

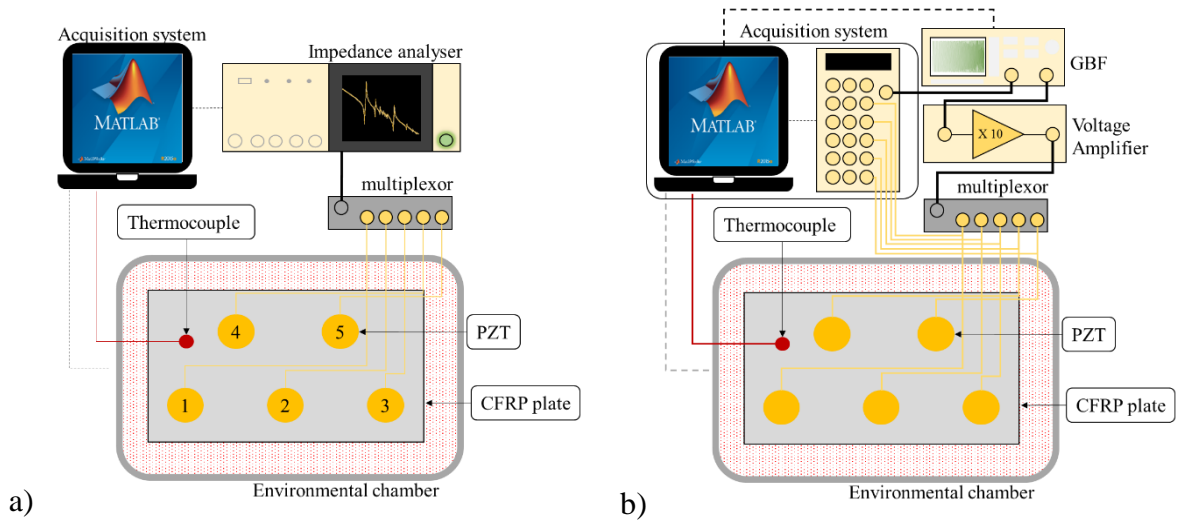

Figure 1 : experimental setup for a) impedance measurements and b) FRF measurements.

For FRF measurements, the exponential sweep signal was emitted with a waveform generator (Agilent 33500 Arbitrary Waveform Generator) and a voltage amplifier (see Figure 1.b). The signal recorded by the PZT sensor was acquired with a Genesis GEN7t card. The emitted signal for each measurement is a $10\ ms$ exponential sweep from $10\ kHz$ to $250\ kHz$ with a sampling frequency of $100\ MHz$. The response signal of the PZT sensor elements was recorded during $15\ ms$ with a sampling frequency of $1\ MHz$ and averaged 100 times to minimize the SNR.

The FRF (see Eq.(5)) was calculated using the Matlab function *tfestimate.m*.

### 3.2 Impedance measurements

*Creation of the database*

The database was realized with capacity measurements on a cycle (increase/decrease) from $10°C$ to $60°C$ by step of $2°C$. It is composed of the parameters of the linear regression described in Eq. (3): the static capacity variation coefficient $\alpha_C$ and the static capacity at $0°C$, $C_s^0$, for each PZT considered. Figure 2 shows the evolution of the static capacity measurements used to create the database for PZT 1. It confirms that the static capacity variation is proportional to the temperature. A hysteresis effect is observed when the temperature variation changes direction. However, this phenomenon will not be taken into account in this article (unlike [14]) but may be studied in future works. Figure 2a indicates that a smaller range of frequencies or even one frequency can be used to determine the temperature variation, but considering a larger scope increases the robustness of the results.

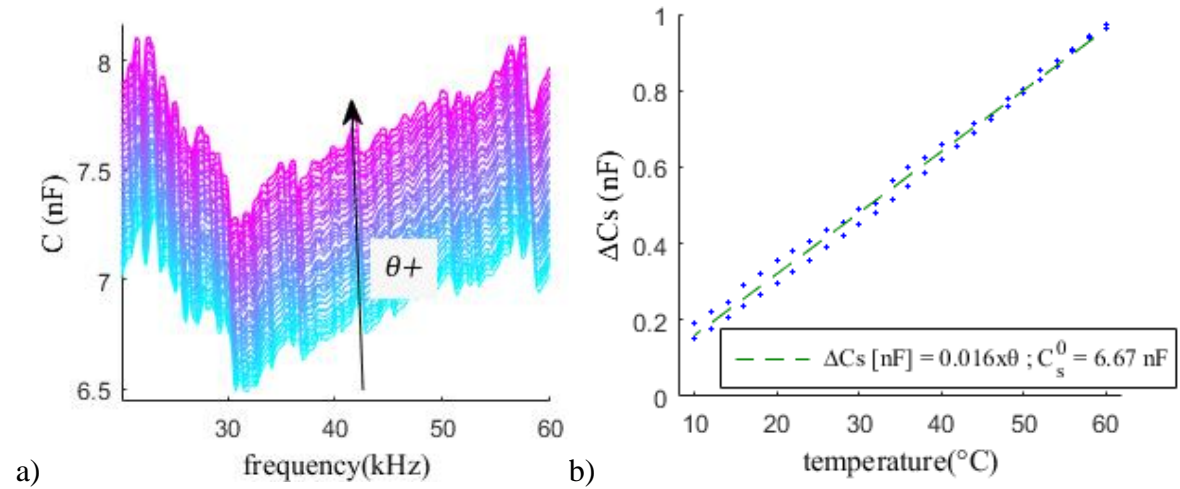

Figure 2 : a) variation of the equivalent capacity described in Eq. (1) of PZT 1 with temperature on the range 20 to 60 kHz, b) variation of the static capacity described in Eq. (2) of PZT 1 on the temperature cycle used as baseline and linear regression computed

|  | PZT1 | PZT 2 | PZT 3 | PZT 4 | PZT 5 |
| --- | --- | --- | --- | --- | --- |
| $\alpha_C$ ($\times 10^{-11} F/°C$) | 1.60 | 1.56 | 1.81 | 1.54 | 1.57 |
| $C_s^0$ ($nF$) | 6.67 | 6.60 | 7.13 | 6.65 | 6.88 |
| $r^2$ | 0.99 | 0.95 | 0.94 | 0.99 | 0.99 |

Table 1: linear regression parameters and coefficient of determination of the capacity variation with temperature for the 5 PZT of the structure under test

The linear regression is estimated (Figure 2b) and stored as a reference in the database: The parameters are slightly different despite the fact that all the PZT are of the same material and are mounted in the same way. This difference can come from PZT health degradation due to previous tests and also differences in the acquisition system (for example, cables are not exactly the same length).

*Temperature estimation*

A second cycle from 0 to $80°C$ and from $80°C$ to $62°C$ by steps of $2°C$ was used as a test sample for temperature estimation. The temperature is estimated using Eq. (4).

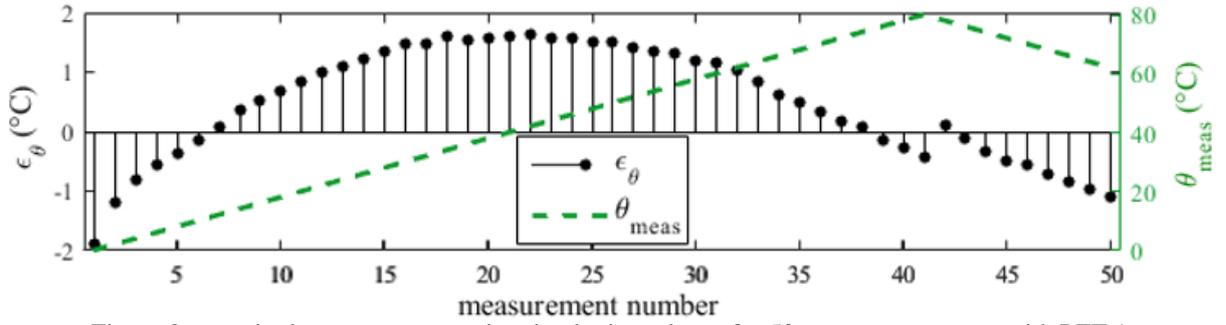
Figure 3: error in the temperature estimation by impedance for 50 test measurements with PZT 1

The error of estimation $\varepsilon_\theta$ (Eq.(9)) indicates a precision of $\pm 2°C$ of the temperature measurement using the static capacity of PZT 1. PZT 2 and 4 had the same precision ($\pm 2°C$), but PZT 3 and 5 showed poorer results (respectively $\pm 10°C$ and $\pm 5°C$). For the moment, experiments have only been made with temperatures from 0 to $80°C$ on one undamaged CFRP plate and showed good results. Further tests on other CFRP structures and with more extreme temperatures will approve the model and delimit its application range. Figure 2a shows a tendency of the capacity to shift under lower frequencies with increasing temperature. This observation might be used for further development to improve the current method.

### 3.3 FRF measurements

*Creation of the database*

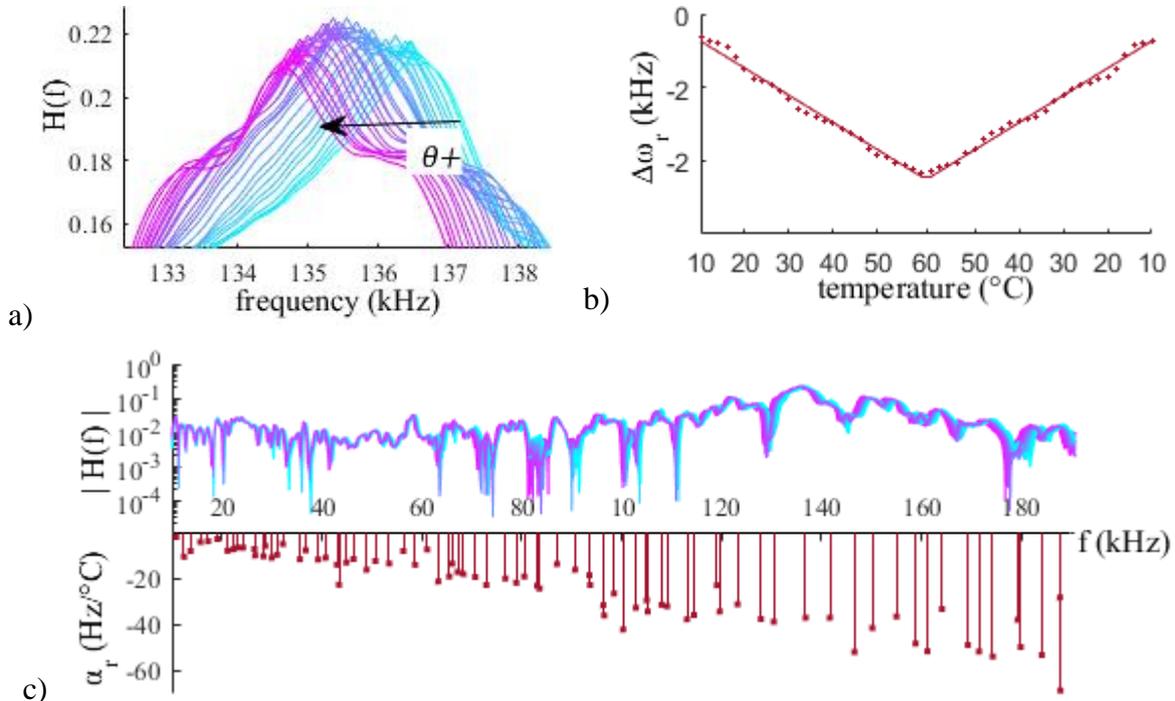

Figure 4: a) MFS of the mode at 136.9 kHz. The triangle represents the position of the peak; b) linear regression applied on the MFS variation of the mode at 136.9kHz ($\alpha_r = -37\ Hz/°C;\ \omega_r^0 = 136.9\ kHz$); c) FRF of path PZT1 - PZT 2 for temperatures varying from 10 to 60°C (top) and thermal drift coefficient for the 80 modes considered (bottom)

The modes are tracked with the Matlab function *findpeaks.m*. An algorithm has been developed in order to follow the MFS that occurs with temperature variations (Figure 4a). This work focuses on the position of the modal pulsation but other parameters like damping and amplitude could be considered in future works. The database was realized with FRF measurements on a cycle (increase/decrease) from $10°C$ to $60°C$ by steps of $2°C$. It is composed of the parameters of the linear regression described in Eq. (6) and (7): the thermal drift coefficient $\alpha_r$ and the pulsation frequency at $0°C$ $\omega_r^0$ for each mode $r$ and for each PZT are considered. Variables $\alpha_r$ and $\omega_r^0$ are determined by studying the MSF of each mode on the cycle used as reference. Figure 4b shows the linear regression applied to the mode at $136.9\ kHz$ in order to determine $\alpha_r$ and $\omega_r^0$. The MSF corresponding to the rising in temperature from $10°C$ to $60°C$ of this same mode is displayed in Figure 4a.

For each mode, the coefficient of determination is computed. It is an important parameter to confirm the validity of the proposed linear regression. The FRF modes are chosen between $10\ kHz$ and $190\ kHz$ (for frequencies higher than $190\ kHz$, the peaks corresponding to mode pulsations are difficult to follow upon temperature variation). Considering this range, 80 modes are detected and their parameters are recorded in the database. The coefficient of determination calculated for the 80 modes are between 0.4 and 1 with a mean of 0.94. This result proves that the linear regression is a good model for the variation of the MSF with temperature. As discussed in the previous section, the authors have decided that no regression model should be applied regarding the MFS variation with frequency. However, Figure 4c shows that the thermal drift coefficient $\alpha_r$ tends to be proportional to frequency for some modes. This observation will be investigated in future works.

*Temperature estimation*

The temperature has been estimated on 12 samples corresponding to FRF measurements at 12 different temperatures between 0 and $30°C$ and 20 modes, using Eq. (8) and the algorithm presented. The error of estimation $\varepsilon_\theta$ (Eq. (9)) indicates a precision of $\pm 2°C$ for each path of the structure. Figure 5 shows results for path PZT 1 – PZT 2.

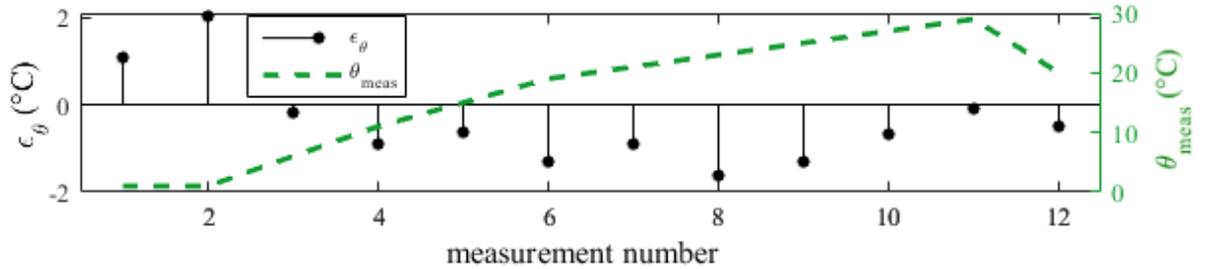

Figure 5 : Error of temperature estimation by MFS for 12 measurements with 20 modes.

The MFS of a mode is qualified by its coefficient of determination, so if the number of modes chosen to determine temperature is lowered, only modes with good coefficients of determination are chosen. The experiments showed that 4 modes are enough to determine the temperature, and actually, the temperature can be estimated with only 1 well-chosen mode, but other environmental parameters (or a damage) can be responsible for the frequency shift of this mode. Using several modes gives more robust temperature estimation because the estimation will be a concatenation of the temperature estimated on each mode.

Experiments were only done on one undamaged CFRP plate. Tests on other plates, with more extreme temperature, and in the presence of damage will be done to study the robustness of this method.

## 4   CONCLUSIONS AND FUTURE WORKS

The temperature has an important influence on SHM and a lot of compensation methods have been proposed in damage diagnosis. However, few studies have considered using the PZT as temperature sensors. In this article, two main axes have been developed to estimate the temperature of a CFRP plate:

- The static capacity, which is proportional to the temperature, permitted to build a database composed of the regression coefficients of the PZTs mounted on the plate. The temperature of a PZT can then be estimated with a precision of $\pm 2°C$.
- The modal frequency-shift (MFS) of the frequency response function obtained with an exponential sweep signal as entry which is also proportional to temperature for most modes. It permitted to build a database with the regression coefficients of each mode for each path between PZT. The temperature of the investigated path could then be estimated with a precision of $\pm 2°C$.

A combination of those two methods could allow one to determine the temperature of a structure on the PZT (static capacity method), and between PZT (MFS method). In an aeronautical context where composite structures are monitored with several PZT and subject to heterogeneous temperature variations, it could give the heat distribution of a structure without any thermometer and improve temperature compensation methods used in damage detection. A fusion of the two methods is also a way to address shortcomings from one method with the other and provide more robust results.

The presented work was realized with homogeneous temperatures on the structure under test. Complementary work is to be done to evaluate the ability of the methods described to detect local temperature variations. Experiments are also planned to check if the temperature estimation using MFS is not altered in presence of a damage.